\documentclass[fleqn,twoside]{article}
\usepackage{espcrc2,epsfig}

\title{Non-perturbative determination of heavy quark action
coefficients
\thanks{We thank RIKEN, Brookhaven National Laboratory and the U.S. Department
of Energy for providing the facilities essential for the completion of
this work.}}

\author{Huey-Wen Lin\address{Department of Physics, Columbia
University, New York, NY, 10027}}
\begin{document}


\begin{abstract}
We propose to determine the coefficients in the Fermilab heavy
quark action by matching the finite-volume, off-shell, gauge-fixed
propagator and vertex functions with those determined in the
exact, relativistic theory.  The matching relativistic amplitudes
may be determined either from short-distance perturbation theory
or from finite-volume step-scaling recursion.  
Tree-level matching and non-perturbative off-shell amplitudes are
presented. \vspace{ -0.3in}
\end{abstract}

\maketitle \vspace{ -0.3in}
\section{Introduction}
The Fermilab action~\cite{Fermilab_action} is the
least restrictive framework permitting lattice methods to be used
for heavy quarks when the quark mass is large compared to the
inverse lattice spacing, $m_{\rm quark}a > 1$. However, at present
the coefficients of this action have only been calculated
using low order lattice perturbation theory. In order to have
better control of errors for the physical quantities calculated on
the lattice, we propose to determine these coefficients
non-perturbatively. We plan to achieve this by imposing a matching
condition between the off-shell quark propagator and quark-gluon
vertex function determined from the Fermilab action, and the same
propagator and vertex computed under more physical conditions.

\begin{figure}[!t]
\vskip +0.0in
\epsfig{figure=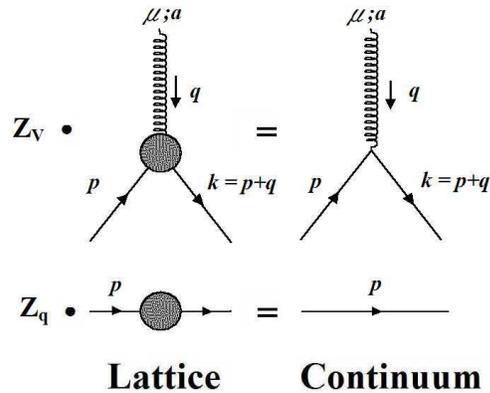,width=\columnwidth}
\vspace{-16mm} \caption{\small\small  Proposal-I.  Match the
lattice result to corresponding continuum perturbation theory
calculation to sufficient order, in the usual RI-MOM scheme. }
\label{PT_matching} \vskip -10mm
\end{figure}
\begin{figure}[!t]
\vskip -0.0in
\epsfig{figure=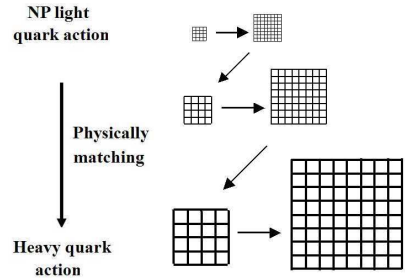,width=\columnwidth}
\vspace{-18mm} \caption{\small\small Proposal-II.
 Match the lattice result, through a step-scaling
technique, to finer non-perturbative, $O(a)$ improved light quark
calculations.} \label{step_scaling} \vskip -7mm
\end{figure}

We propose two ways to achieve this goal.
 The most naive method would be matching the propagator and
vertex function computed in the continuum theory up to some order
in perturbation theory with the RI-MOM NPR~\cite{RI_NPR}
calculation on the lattice (Figure~\ref{PT_matching}). This would
be done through the ``matching window''~\cite{hwlin02}:
$\Lambda_{QCD}^2 \ll m^2-p^2 \ll 1/a^2$ .
The alterative will be purely non-perturbative, by connecting the
original calculation with one performed at a smaller lattice
spacing $a^\prime = a*\epsilon$ (as shown in
Figure~\ref{step_scaling}).
A further advantage of such an entirely non-perturbative matching
is that the kinematic ``window'' for performing such matching now
has no lower bound since there is no requirement that the
perturbation theory be accurate.

\vspace{-0.2in}
\section {Off-shell Fermilab action}
\vspace{-0.1in}
 Extending the off-shell $O(a)$ improvement of light
quarks~\cite{Martinelli_offShell} to heavy quark, we will adopt
the approach of using an unchanged fermion action in the
simulation: \vskip -0.2in
\begin{eqnarray}
&&\hskip -0.2in S  =  \sum_n \bar{\psi_n}(D_{FL} + m)\psi_{n}
\nonumber
\end{eqnarray}
\vskip -0.1in
 \hskip -0.15in
and improved quark fields of the form: \vskip -0.2in
\begin{eqnarray}
&&\hskip -0.2in \psi \rightarrow \hat{\psi} = \mathcal{Z}_{q}^{-
\frac {1}{2}}
[1+a{c_q}' (D_{FL} + m)\nonumber\\
&&\hskip 0.8in - ac_{NGI}^t \gamma_0 \partial_0 + ac_{NGI}^s
\gamma_i
\partial_i
]\psi\nonumber\\
&&\hskip -0.2in \bar{\psi} \rightarrow \hat{\bar{\psi}} =
\mathcal{Z}_{q}^{- \frac {1}{2}} \bar{\psi} [1+a{c_q}'
(-D_{FL}^{\leftarrow} +
m)\nonumber\\
&&\hskip 0.8in  - ac_{NGI} ^t \gamma_0 \partial_0^{\leftarrow}-
ac_{NGI}^s \gamma_i
\partial_i^{\leftarrow}] .\nonumber
\end{eqnarray}
\vskip -0.05in
\hskip -0.15in%
where $D_{FL}$ is the Fermilab Dirac operator. Note:
$\mathcal{Z}_{q}$ and c's are all functions of $ma$. In summary,
we have eight parameters as a function of $ma$ to be determined:
$\zeta$, $r_s$, $c_E$, $c_B$ (in the action) and
$\mathcal{Z}_{q}$, $ {c_q}'$, $c_{NGI}^t$ and $c_{NGI}^s$ (in the
improved quark fields).

%
If we look at the low spatial momentum behavior of the Dirac
structure we find that there are more than the six required
conditions from $\Lambda_{\mu}$ and the two from $\hat{S}(p)$ ,
and thus we are able to fix these eight parameters.
\section {Tree-level coefficient calculations}
\vskip -0.1in
Before we start the complicated computer calculations, we can test
our proposal-I by starting with lattice perturbation theory in
tree-level and matching to
the corresponding continuum 
quantities:
\begin{eqnarray}
 &&\hskip -0.2in\{\Lambda_{\mu}(p,q;\mu)|_{m^2-p^2=\mu^2} \}^{phys}\nonumber \\
  &&= \mathcal{Z}_q(\mu;a)^{-1}\mathcal{Z}_3^{1/2}(\mu;a)\Lambda_{\mu}^{\rm
lat}(p,q;a)|_{m^2-p^2=\mu^2}
       \nonumber
\end{eqnarray}
\vskip -0.3in
\begin{eqnarray}
&&\hskip -0.2in\{\langle \hat{S}(p) \rangle^{-1}|_{m^2-p^{2} =
{\mu}^{2}}\}^{phys}\nonumber \\
  &&= i \mathcal{Z}_q [ip_0 \gamma_0 A_0^{\rm lat}(p^2) +i p_i\gamma_i A_i^{\rm
lat}(p^2)]+ B^{\rm lat}(p^2)
        \nonumber
\end{eqnarray}
\vskip -0.3in
\begin{eqnarray}
&&\hskip -0.2in  \mathcal{Z}_3 D(q^2;\mu)|_{q^{2}={\mu}^{2}}
=\frac{1}{\mu^2}\nonumber
\end{eqnarray}
\vskip -0.1in
Expanding in the variables $p_ia$ and $q_{\mu}a$ up to $O(a)$ we
determine:
\begin{eqnarray}
&&\hskip -0.2in \zeta = \cos(p_0a+q_0a/2) = \cos (p_0a) - q_0a/2
\sin(p_0a)
\nonumber\\
&&\hskip -0.2in \mathcal{Z}_q^{-1}  = (1+4c_qma){\cos
(p_0a+q_0a/2)}
\nonumber\\
&&\hskip -0.2in c_{NGI} ^t= \frac{1}{2ma}
\hskip+0.05in\{\frac{p_0a}{\sin
p_0a}\nonumber\\
&&\hskip 0.6in\times(\cos (p_0a+q_0a/2)-1)(1+4c_qma)
  \}\nonumber\\
&&\hskip -0.2in c_{NGI} ^s = 0 \nonumber\\
&&\hskip -0.2in c_q= 1/2 - c_{NGI}^t \cos (p_0a+q_0a/2) \nonumber\\
&&\hskip -0.2in c_E  = 2c_q
\nonumber\\
&&\hskip -0.2in c_B = 2c_q\zeta
 \nonumber\\
&&\hskip -0.2in r_s = 2c_q\zeta \nonumber
\end{eqnarray}
Taking the $p_{0}a$ and $ma$ $\ll$ 1 limit, we reproduce the known
coefficients
\\ {}{}$c_E$ = $c_B$ = $\zeta$ = $r_s$ = 1,
$c_{NGI}^{\{t,s\}}$ = 0 ,\\$c_q$ = 1/2, $\mathcal{Z}_q$ = 1 $-$ 2$ma$ \\
for the improved, light quark action.
%
\section {Non-perturbative Green's functions}
\vspace{-0.1in}
We have studied connected 2-point Green's functions (quark and
gluon propagators)~\cite{hwlin_future} and have produced answers
consistent with published papers.
The accuracy achievable with moderate statistics  is adequate to
determine those improvement coefficients that are constrained by the
2-point functions. From now on, we will focus on the least known
quantity: the quark-gluon vertex function.


The quark-gluon vertex function can be calculated from the
following 3-point function~\cite{qgv}:
\begin{eqnarray}
&&\hskip -0.2in V_{\mu}^a(x,y,z)_{\alpha, \beta}^{ij}
= \langle S_{\alpha, \beta}^{ij}(x,z) A_{\mu}^a(y)\rangle.
 \nonumber 
\end{eqnarray}
The Fourier transformed
amputated vertex function in momentum space is
\begin{eqnarray}
&&\hskip -0.2in \Lambda_{\mu}^{
a,{\rm lat}}(p,q)_{\alpha,\beta}^{ij}\nonumber\\
&&\hskip 0in =
\langle \hat{S}(p)\rangle^{-1}V_{\nu}^a(p,q)_{\alpha,
\beta}^{ij}\langle\hat{S}(p+q)\rangle^{-1} \langle
D(q)_{\mu\nu}\rangle^{-1}\nonumber
\end{eqnarray}
We calculate the quantity
$  \frac{1}{4}Im \sum_{\mu}Tr(\gamma_{\mu}\Lambda_{\mu, R}^{ {\rm
lat}})$ in the case where q = 0.

As a first step we plan to non-perturbatively determine the
coefficients in light quark clover action, since only four
parameters are involved: $c_{sw}$ (in the action) and
$\mathcal{Z}_{q}$, $ {c_q}'$, and $c_{NGI}$(in the improved quark
fields). Our gauge configurations use the Wilson gauge action with
$\beta$ = 6.0 , $a^{-1}$= 1.922 GeV, $16^3\times$32 lattice size.

\begin{figure}[!t]
\vskip -0.3in
\includegraphics*[width=\columnwidth]{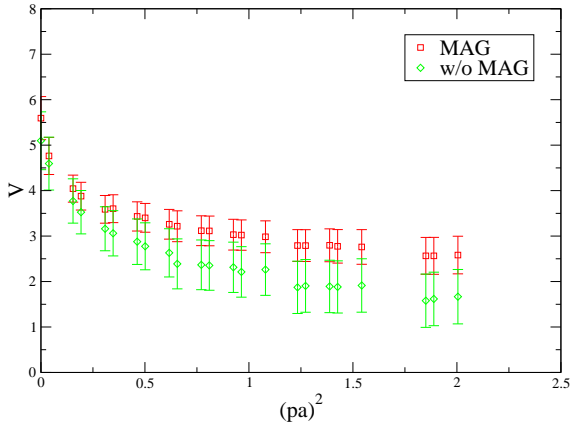}
\vskip -10mm
 \caption{\small Gluon ``q = 0'' vertex function
 calculation from the tadpole improved clover action
with $\kappa=0.137$ from 43 configurations.
} \label{asym_ver_MAG_comp}%
\vskip -7mm
\end{figure}

We use Landau gauge and proceed by two approaches: 1.\ fix to
maximal axial gauge (MAG) before Landau gauge fixing; 2.\ fix only
to Landau gauge. We found that the two different procedures agree
for the gluon propagator, have slight effects on quark propagators
(decreasing as the momentum increases) but give very different
results in the vertex calculation~\cite{hwlin_future}, as shown in
Figure~\ref{asym_ver_MAG_comp}. We believe the differences are
caused by Gribov copies. From now on, we will follow procedure 2.
for better control over which gauge orbit is selected by the
Landau gauge fixing procedure, and for the cleaner signal.

\begin{figure}[!t]
\vskip -0.3in
\epsfig{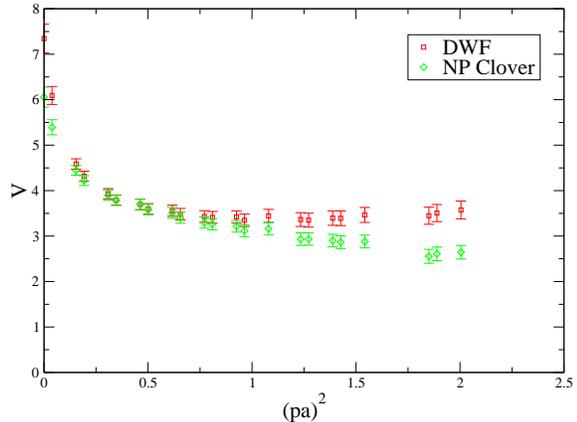}
\vspace{-15mm} \caption{\small Gluon vertex function with ``q =
0'' kinematics for 379 configurations.  } \label{cloverDwfAsym}
\vskip -7mm
\end{figure}

 Figure~\ref{cloverDwfAsym} shows the comparison of
the unrenormalized q = 0 vertex functions from NP clover with
$\kappa$ = 0.13445 without off-shell improvement, and with domain
wall fermions
 ($L_S$ = 16, $M_5$ = 1.8). DWF is automatically $O(a)$ off-shell
improved and the vertex nicely becomes independent of momenta in
high momentum limit, as one would expect from a perturbative
analysis. It is precisely this difference in behavior that will
enable us to determine the coefficients to off-shell improve the
clover action, first in the light quark limit, and later as a
function of $ma$. The masses of the pseudoscalar meson calculated
from these two actions are similar, around 500 MeV. Once we impose
the $O(a)$ off-shell improved quark field clover action, the
difference should be reduced.

The ``q = 0'' kinematics has cleaner signal than ``q $\ne$ 0'' and
would be a better candidate for a step-scaling proposal,
%
%
while the ``q $\ne$ 0'' would be required to implement
proposal-I~\cite{hwlin_future}. This would require better control
of the statistical errors. The large error bar at ``q $\ne$ 0''
comes mainly from the weak correlations between the quark
propagator and the gluon fields. We are currently working on using
sources with definite momentum to achieve improved statistical
accuracy.

%


\section{Summary and Outlook }
\vspace{-0.1in}
 In this paper, we have proposed two methods  for
non-perturbatively determining the coefficients of the heavy quark
action. We calculated the tree-level coefficients and reproduced
the known light quark limit. It seems possible to determine the
Fermilab coefficients with sufficient effort through either of our
proposals. We will next look at other Dirac projections and
momentum configurations. Once we have investigated the optimal
method more thoroughly, we will start with non-perturbative
determination of $c_{sw}$ for the light quark action and reproduce
the Alpha collaboration result. We will then move on to the large
quark mass case: the Fermilab action.

\vspace{-0.05in}
\section*{ACKNOWLEDGMENTS}
I thank N. Christ and P. Boyle for physics discussions, C. Dawson
for comparing RI-MOM quark propagator data, J. Skullerud for a
vertex function calculation comparison, T. Bhattacharya for
comments in the lattice conference and the RBC collaboration for
the physics system and QCDSP machine time.

\vspace{-0.1in}

\end{document}